\documentstyle[11pt,aaspp4]{article}

\slugcomment{Accepted for publication in Part 1 of The Astrophysical Journal.}

\lefthead{P. Lanoix, G. Paturel and R. Garnier}
\righthead{Bias in the Cepheid PL relation }

\begin{document}
\title{Bias in the Cepheid PL relation \\}

\author{P. Lanoix\altaffilmark{1}, G. Paturel and R. Garnier}
\affil{CRAL (UMR 5574 CNRS) - Observatoire de Lyon \\
        F69230 Saint-Genis Laval, FRANCE}
\altaffiltext{1}{Universit\'e Claude Bernard Lyon I F69622 Villeurbanne, FRANCE}
\authoremail{Lanoix, Paturel, Garnier@obs.univ-lyon1.fr}

\begin{abstract}
We show that the  Cepheid PL relation is affected by a Malmquist
type bias, the so-called {\it population incompleteness bias}.
Its calculated slope appears shallower than the true one because of
the cutoff in apparent magnitude resulting from the instrumental limiting
magnitude. Furthermore, the use of the PL relation, even with
the correct slope, leads to underestimation of distances which is not negligible.
We confirm this finding by studying simulated PL relations, and we show
that this bias may be as large as 0.2 or 0.3 magnitudes on distance modulus.
We also test the efficiency of a cutoff in $\log P$ and show that it is a good
way to minimize this bias.
However, a correction of this is difficult as long as the completeness of the
sample is not perfectly well established.
\end{abstract}


\section{Introduction}
The Cepheid Period-Luminosity relation (hereafter PL relation)
is one of the major tools used for computing
extragalactic distances from data like those provided by the 
{\it Hubble Space Telescope} (HST).\\
However, Teerikorpi (\cite{tee84}) has pointed out for a long time that 
such a linear relationship is prone to be affected by biases.
For extragalactic Cepheids a bias is expected, identical to that encountered
in a cluster of galaxies, the so-called {\it population incompleteness
bias} (\cite{tee87,bot87}).\\
Sandage (\cite{san88}) has already noticed that truncating a complete
sample of Cepheids in LMC leads to too shallow a slope of the PL relation.
We intend to show that HST observations, although much deeper in apparent
magnitude, are affected by this kind of bias. This is because
a sample of Cepheids in a  given external galaxy (thus, at the same
distance) is complete only at the bright end of the
Cepheid luminosity function (i.e. only for large $\log P$ values), since 
observations are limited to a given apparent limiting magnitude. \\
After introducing the data and the method used (section 2),
we investigate how such a bias affects the slope, $a$, (section 3), by
studying NGC 4536 as an example. We then highlight
the existence of the bias in this way. In section 4
we analyze the effect on the zero-point $b$, using the 
correct slope of the PL-relation.
Section 5 presents a simulation that confirms previous results and
gives an estimate of the bias,  
as well as its reduction by applying a cutoff in $\log P$ .\\
It is important to address this question because its effect is poorly 
studied at the present time, and, yet, is larger than generally admitted.

\section{Data and distance moduli}
We use a sample of 750 Cepheids which have both V and I 
photometry available from literature. These Cepheids are located 
in 23 galaxies, 14 of them have Cepheids
observed with HST whereas the other 9 galaxies have 
only ground based observations. We checked all light curves in order
to detect overtone pulsators (low amplitude, $\log P < 1$ and
symmetrical curves) and to remove them from our sample since they
are subject to a different PL relation (we didn't try to correct their period).  \\
In order to compute the distance moduli of these galaxies, we choose
the absolute calibration of the PL relation from Gieren et al. (\cite{gie98}).
Their PL calibration is based on infrared Barnes-Evans surface 
brightness technique (note that it is insensitive to both 
Cepheid metallicity and
reddening). We then avoid comparing PL relations of different galaxies 
with different metallicities and inaccurate reddenings. The PL 
relations are :\\
\begin{equation}
M_{V} = -2.769 \log P - 1.294 \\
\label{plrv}
\end{equation}
\begin{equation}
M_{I} = -3.041 \log P -  1.726
\label{plri}
\end{equation}
We can compute easily the V and I apparent distance moduli for
each Cepheid :
\begin{equation}
\mu_{V} = V - M_{V} 
\end{equation}  
\begin{equation}
\mu_{I} = I - M_{I} 
\end{equation}  
The supposed true distance moduli are then :
\begin{equation}
\mu_{0} = \mu_{V} - R (\mu_{V} - \mu_{I})
\label{mu}
\end{equation}
where 
\begin{equation}
R = \frac{A_{V}}{A_{V}-A_{I}} = 2.446
\end{equation}
according to \cite{car89}, $A_{V}$ and $A_{I}$ being respectively
the V and I extinction coefficients.
The supposed true distance modulus of a galaxy is then assumed to be the mean of
the individual distance moduli of its Cepheids. 

\section{Effect on the PL relation slope}
The V-band slope is very well defined from ground-based 
observations of LMC Cepheids. Gieren et al. (\cite{gie98}) found $a=-2.769\pm0.073$.
Tanvir (\cite{tan98}) found $a=-2.774\pm0.083$ and
Madore \& Freedman (\cite{mad91}) found $a=-2.76\pm0.11$. 
We will adopt :
\begin{equation}
a = -2.77 \pm 0.09
\end{equation}
The slope of the PL relation cannot be determined from distant galaxies.
We almost always obtain apparent 
slopes ($a'$) significantly shallower than
the LMC one (assumed to be the true one).
Yet, the slope of the PL relation does not vary with 
metallicity. This is predicted by models (\cite{chi93})
and is widely acknowledged even by people who claim that metallicity effects are
strong (\cite{bea97}). Therefore, we cannot argue about metallicity effects.\\
The low value of $a'$ can actually be traced to 
the population incompleteness
bias resulting from apparent limiting magnitudes ($V_{lim} \approx 26.5$ 
for HST measurements and $V_{lim} \approx 21.5$ for ground-based observations).
\placefigure{fig1}
\placefigure{fig2}
Figure \ref{fig1} and \ref{fig2} show respectively the 
V and I PL relations for one of the 
best measured galaxies, NGC 4536, and illustrate how the bias works.
For instrumental reasons, apparent magnitudes  are limited to a given
limit $V_{lim}$. Thus, the distribution in the plot $V$
against $\log P$ (or similarly $M_V$ against
$\log P$ for a given distance modulus) is distorted. If we force a linear
regression, the line will tend to pass by the point having the largest 
inertia (note $A$ in fig. \ref{fig1} and \ref{fig2}).
This leads to a shallower slope. 
Assuming a Gaussian distribution of residuals (as a first approximation of 
top-hat distribution), the model given 
by Paturel et al. (\cite{pat97})
predicts a slope $a'$ :
\begin{equation}
a'=\frac{a}{1+ \sigma / (V_{lim} - \mu  - M_{Max})}
\label{slope}
\end{equation}
where $a$ is the unbiased slope, $\sigma$ is the scatter
of the V band PL relation in magnitudes ($\sigma \approx 0.3$ mag.),
$\mu$ is the adopted distance modulus
and $M_{Max}$ the brightest end of absolute V magnitudes ($M_{Max}
\approx -7$). 
\placefigure{fig3}
Figure \ref{fig3} gives the predicted variation of $a'$ with the distance 
modulus for both ground-based and HST observations. On the same
figure we give the observed slopes obtained by a direct regression for
Cepheids in 23 galaxies (ground-based and HST observations are 
represented with different symbols). The agreement fully supports
the interpretation of the bias and proves in this 
way that PL relations in external galaxies are definitely affected by
incompleteness bias. It must be noted that the
slope dramatically increases with distance modulus, preventing
us from using an observed slope without caution. \\
The inverse regression line is a way which should
lead to the correct result (\cite{kel96}). 
Unfortunatelly, it is subject to the opposite bias (bias in $\log P$) because 
very long period Cepheids are difficult to detect.
Furthermore, the inverse slope varies with the scatter of the sample
prohibiting the direct comparison with calibrating PL relation whose
scatter is smaller. This situation is identical to that encountered
for the inverse Tully-Fisher relation (see \cite{tee90}).

\section{Effect on the zero-point}
We would like to stress the most important consequence
of the bias. Even with the right slope, the determination of the 
distance modulus will be underestimated when data are limited
in apparent magnitude.
\placefigure{fig4}
This fact can easily be understood from the distribution of the
points in figures \ref{fig1} and \ref{fig2}. Because of the cutoff 
in magnitude there are more points above the mean PL line for
short periods. Of course, since the residuals on $\mu_{V}$ and $\mu_{I}$
are correlated, the effect of incompleteness is reduced in the second part
($\mu_{V} - \mu_{I}$) of the equation (\ref{mu}). However, the effect on the first
part ($\mu_{V}$) of this equation remains affected by incompleteness, and
this term is moreover one order larger than the second one 
($R(\mu_{V} - \mu_{I}) / \mu_{V} < 5 \% $).\\
In order to highlight this effect,
the mean distance modulus is calculated for NGC 4536 by
progressively removing the shortest period Cepheids.
Figure \ref{fig4} shows the variation of the mean distance modulus with 
the lower $\log P$ cutoff ($\log P_{l}$). The first point (on the left) 
represents the mean distance modulus
of NGC 4536 when using its 73 Cepheids {\it vs.} the lower $\log P$ of these 73
Cepheids. The next point represents the mean
distance modulus {\it vs.} the lower $\log P$ of the 
72 Cepheids having the larger periods. The same applies to 71 Cepheids until
1 Cepheid.
The effect on the mean distance modulus as a function of period cutoff
is then directly readable on this figure.  Presently, the 
detection of a plateau in the growth curve of mean distance
modulus seems the most efficient
way to get unbiased distance moduli from Cepheids PL relation.
The beginning of the plateau indicates the cutoff in $\log P$
which should be applied ($\log P_{l} = 1.25$ in this case). \\
In this example, the biased modulus equals to 30.50 while the
unbiased modulus is 30.76. 
This effect is therefore, far from being negligible.
The rule of thumb for calculating unbiased distance moduli
is simply to use only $\log P$ values larger than the given limit $\log P_{l}$. 
This limit can also be calculated from figure \ref{fig1}:
\begin{equation}
\log P_{l}= \frac{ V_{lim} - \mu  - b - 2\sigma}{a}
\label{cut}
\end{equation}
For NGC 4536, using the PL relation given in equation (\ref{plrv}),
and the same parameters as in 
relation (\ref{slope}), $V_{lim}=26.5$ and
$\sigma=0.3$, we obtain $\log P_{l}=1.19$ for $\mu=30.50$. 
From the growth curve of distance modulus (fig. \ref{fig4})
it appears that this limit is quite acceptable. The definition of
$V_{lim}$ is namely crucial and difficult.

\section{Simulation}
\subsection{Construction of simulated PL relations}
In order to confirm that our interpretation of the incompleteness bias
is the right one, we construct simulated PL relations in the following way :
for each Cepheid of a simulated galaxy, we first choose a true 
distance modulus ($\mu_{True}$) and we determine randomly
according to a Gaussian shape the following values :
\begin{itemize}
\item $\log P$ ; ($<\log P> = 1.0, \sigma = 0.35$)
\item $E_{B-V} = (B-V) - (B-V)_{0}$ ; ($<E_{B-V}>= 0.08, \sigma = 0.04$)
\end{itemize}
Using equations (\ref{plrv}) and (\ref{plri}) we compute the V and I 
absolute magnitudes of
the Cepheid\,\footnote[1]{note that the color distribution 
$(V-I)_{0} = 0.272 \log P + 0.432$ is underlying} and add to them
 a random dispersion $\Delta$, intrinsic dispersion
of PL relations with $<\Delta>=0$ and $\sigma_{\Delta} = 0.3$. 
We then compute and add V and I absorptions according to $A_{V} = 3.3 E_{B-V}$
and $A_{I} = 1.95 E_{B-V}$. We add random error measurements, $\epsilon$,
calculated as a linear function of the distance modulus,
$\epsilon = 0.2 -(32-\mu_{True})/45$, with no correlation between V and I, and
obtain : 
\begin{equation}
V = \mu_{True} -2.769 \log P -1.294 + \Delta + A_{V} + \epsilon_{V} \\
\label{magobsv}
\end{equation}
\begin{equation}
I = \mu_{True} -3.041 \log P -1.726 + \Delta + A_{I} + \epsilon_{I} 
\label{magobsi}
\end{equation}
Finally, we compute the probability for the Cepheid to be detected both in
the V and I bands. We draw a random parameter $t \in [0, 1]$ and compute 
the quantity :
\begin{equation}
t_{0} = \frac{1}{1+e^{\alpha(V-V_{lim})}} 
\end{equation}
Whenever $t \le t_{0}$, the Cepheid may be observed by the HST and we keep
it in our simulated galaxy. In the other case it will be rejected. 
We assume $\alpha = 3$ and $V_{lim} = 26.5$ for HST measurements. \\
We also assume that a typical galaxy has 250 Cepheids, so that about
80 may be detected for $\mu_{True} \approx 30.5 $

\subsection{Results}
We first simulate a galaxy similar to NGC 4536 by locating it at a true
distance modulus $\mu_{True}=30.76$. Figures \ref{fig5} and
\ref{fig6} show respectively the kind of absolute V and I PL relations we obtain. 
\placefigure{fig5}
\placefigure{fig6}
This simulated galaxy has for instance 77 observable Cepheids and is comparable in
every respect to NGC 4536. We can reduce these data to
compute its distance modulus ($\mu_{0}$) according to the procedure 
described in section 2.\\
We find :
\begin{equation}
\mu_{0}= 30.59
\end{equation}
And then the bias is :
\begin{equation}
\mu_{0} - \mu_{True}= -0.17
\end{equation}
At the very least, when using HST data, authors involved in the HST Key
Project make a cut at $\log P =1$. 
When we choose such a lower limit in our simulation, the bias induced
is worth $-0.12$ mag.
\placefigure{fig7}
Figure \ref{fig7} shows the variation of its computed distance modulus as a
function of $\log P_{l}$. The behaviour is the same as the one observed
for NGC 4536 (see fig. \ref{fig4}) and, as such, it confirms our 
interpretation. We plot the plateau on figure \ref{fig7}, which is obtained with
a cutoff at $\log P_{l} = 1.24$ (then $<\log P>=1.45$). \\
If we compute the theoretical cutoff according to equation (\ref{cut}), we
obtain $\log P_{l}=1.23$ in close agreement with the real value. 
This formula (\ref{cut}) may be very useful whenever the indentification of the plateau is
difficult.\\
Moreover, in order to make sure that the result does not depend
 on this particular
random draw, we prepare a random set of 200 galaxies and we place them
at increasing distance moduli.\\
We first test the variation of the slope of the PL relation :
figure \ref{fig8} shows the variation
of the slope as a function of $\mu_{True}$. This figure is directly 
comparable to figure \ref{fig3} and fully supports our interpretation. \\
\placefigure{fig8}
Of greater significance is our evaluation of the bias on the mean distance moduli $<\mu_{0}>$ 
from the same set of galaxies as a function of
$\mu_{True}$ when the correct slope is used.
\placefigure{fig9}
One can see that the bias is indeed far from negligible (fig. \ref{fig9}) .
It reaches :
\begin{equation}
<\mu_{0}>-\mu_{True} = 0.20 \pm 0.01 (\sigma = 0.20)
\end{equation}
for $\mu_{True}=32.2$ and $V_{lim}=26.5$. \\
We also test the efficiency of the bias correction by applying a $\log P$
cutoff. We use our model (see eq. [\ref{cut}]) and compute again 
$<\mu_{0}>-~\mu_{True}$ (open dots in fig. \ref{fig9}).
The importance
of the bias is strongly reduced in this manner although it still
exists. Clearly, the bias depends critically on the limiting $V_{lim}$.
Our simulation is made with a constant $V_{lim}=26.5$, while, e.g. in the Key
Project conducted with the HST, distant galaxies have observations
taken with longer exposure times leading to fainter $V_{lim}$ (for
galaxies in Virgo and Fornax clusters, the limiting magnitude may be between 
26.5 and 27 mag). This will infer different biases depending on the
actual limiting magnitude.\\
From a practical standpoint, one way of taking this bias 
into account might be to cut 
systematically the studied sample in $\log P$ according to equation
(\ref{cut}). For instance, the effects on the distance moduli of
NGC 4321 and NGC 2541 are respectively 0.09 and 0.05 magnitudes.
Another approach, more statistically oriented, could be to use values given in figure
\ref{fig9} and to add the value of the bias we compute to the distance
modulus obtained from the complete sample.

\section{Conclusion}
In conclusion, we draw the attention to the
difficulties involved in using the PL relation. The slope cannot be determined without
considering the population incompleteness bias. 
Furthermore, it is most important to note that the use of the 
relation, even with the correct slope, should at least
be limited to long period Cepheids. The correction for
bias effect is difficult as long as the completeness of the
sample is not perfectly well established. We consider, however, 
that this incompleteness bias should be systematically taken into account when 
deriving extragalactic distances from Cepheid PL relation, for instance
according to the equation (\ref{cut}), even though
it may be of little importance in some cases.

\newpage

\acknowledgments
We would like to thank Joseph Necker Lanoix for his useful corrections, 
as well as the anonymous referee who helped us to improve the paper.

\clearpage

\figcaption[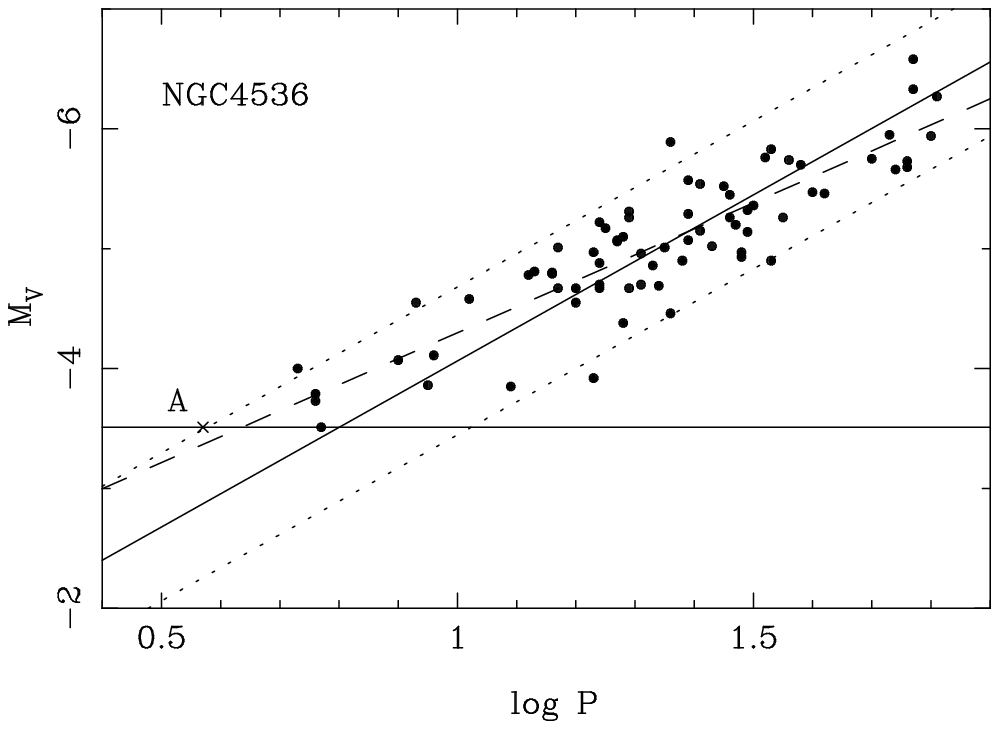]{See next figure. \label{fig1}}
\figcaption[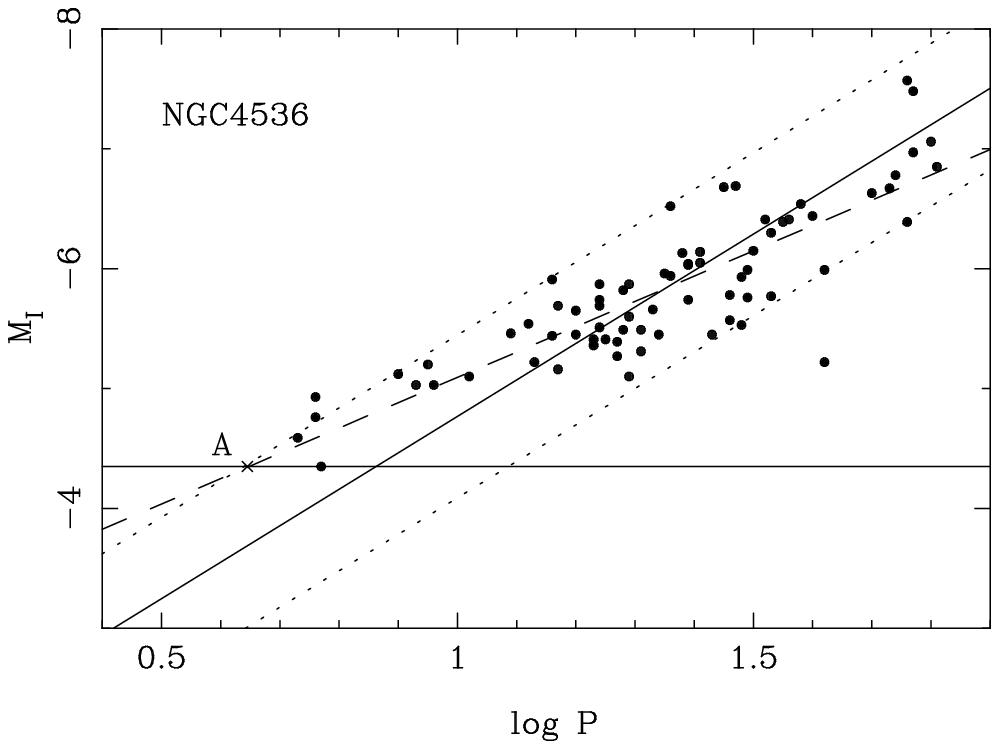]{
NGC 4536 V and I band absolute PL relations. The absolute magnitudes are simply
$M_{V} = V - <\mu_{V}>$ and $M_{I} = I - <\mu_{I}>$, $<\mu_{V}>$ and
$<\mu_{I}>$ being respectively the mean V and I apparent distance
modulus of the considered galaxy.
The solid line corresponds to the adopted PL relation (eq. [\ref{plrv}] and [\ref{plri}]).
The dotted lines give the corresponding $\pm 2 \sigma$ limits.
The dashed line gives the relation calculated from a direct regression.
The horizontal line gives the approximate
cutoff in absolute magnitude resulting from the instrumental
limiting magnitude. The linear regression (dashed line)
forced onto the distorted distribution tends to pass through
point $A$, leading to too shallow a slope. \label{fig2}}

\figcaption[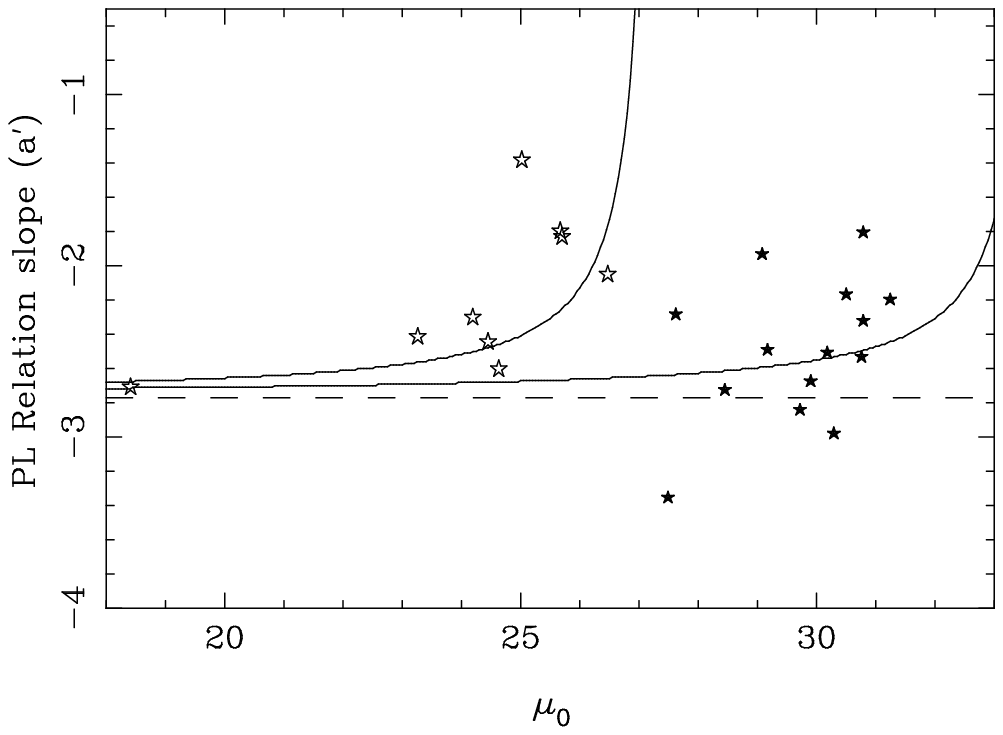]{
Apparent slope $a'$ of the PL relation as a function of
the distance modulus. The trend is typical of a statistical bias. Open stars
represent ground-based observations whereas filled stars
correspond to HST observations.
The solid curves give the predicted variation in both cases from our
model (see eq. [\ref{slope}]). \label{fig3}}

\figcaption[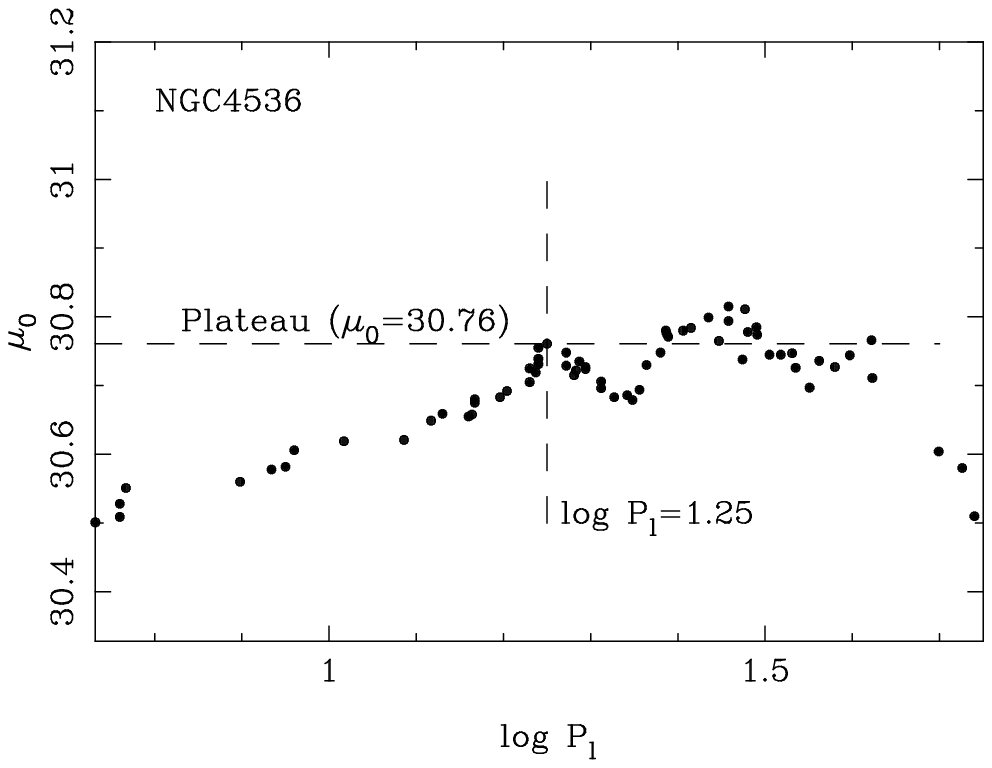]{
Variation of the mean distance modulus {\it vs.} $\log P_{l}$ for NGC 4536.
The mean distance modulus depends on the adopted cutoff
in $\log P$ (see text). \label{fig4}}

\figcaption[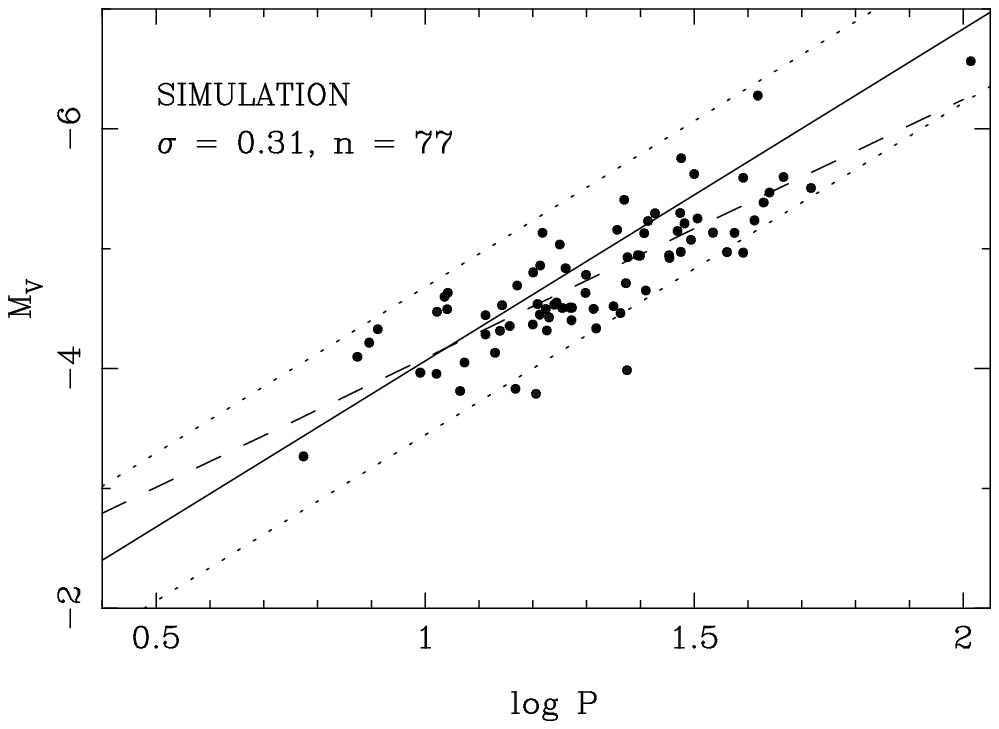]{See next Figure. \label{fig5}}

\figcaption[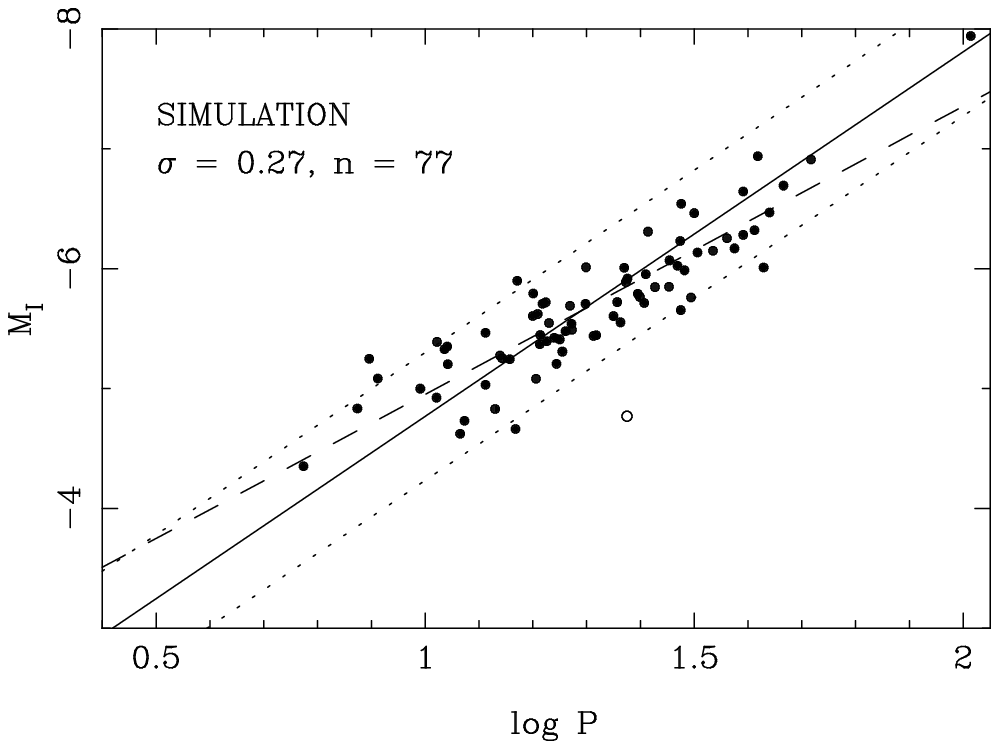]{ 
Simulation of V and I band absolute PL relations. The simulated galaxy
is located at $\mu_{True} = 30.76$.
The solid line corresponds to the adopted PL relation (see eq. [\ref{plri}] and [\ref{plrv}]).
The dotted lines give the corresponding $\pm 2 \sigma$ limits.
The dashed line gives the relation calculated from a direct regression.
\label{fig6}}

\figcaption[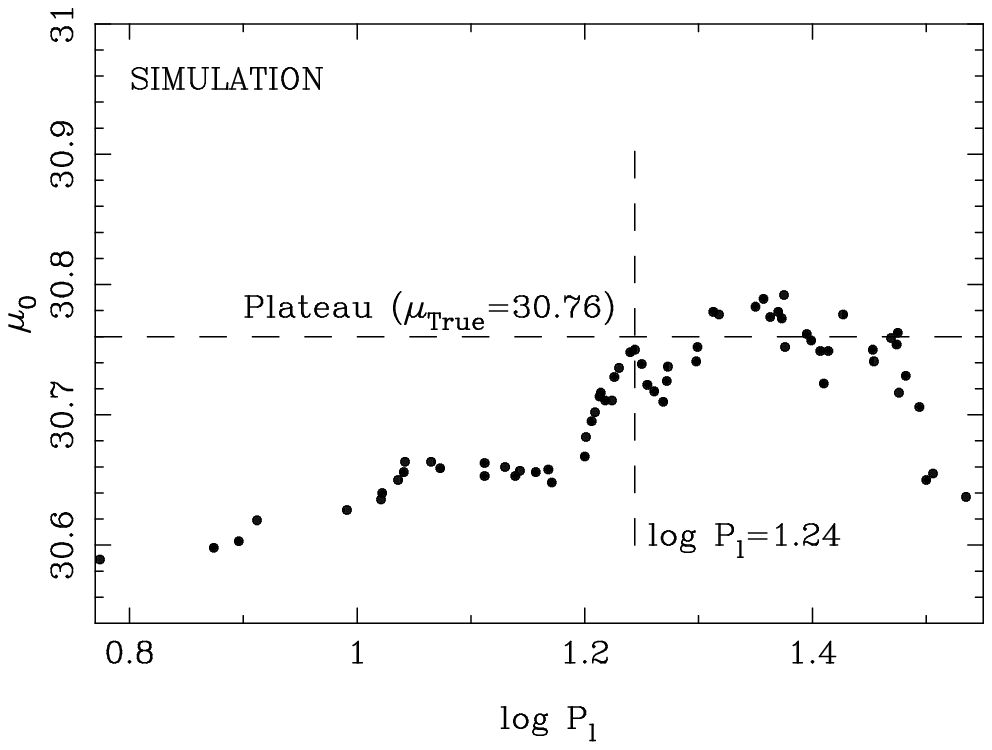]{
Variation of the mean distance modulus {\it vs.} $\log P_{l}$ for a typical
simulated galaxy located at $\mu_{True} = 30.76$.
The mean distance modulus depends on the adopted cutoff
in $\log P$ (see text). \label{fig7}}

\figcaption[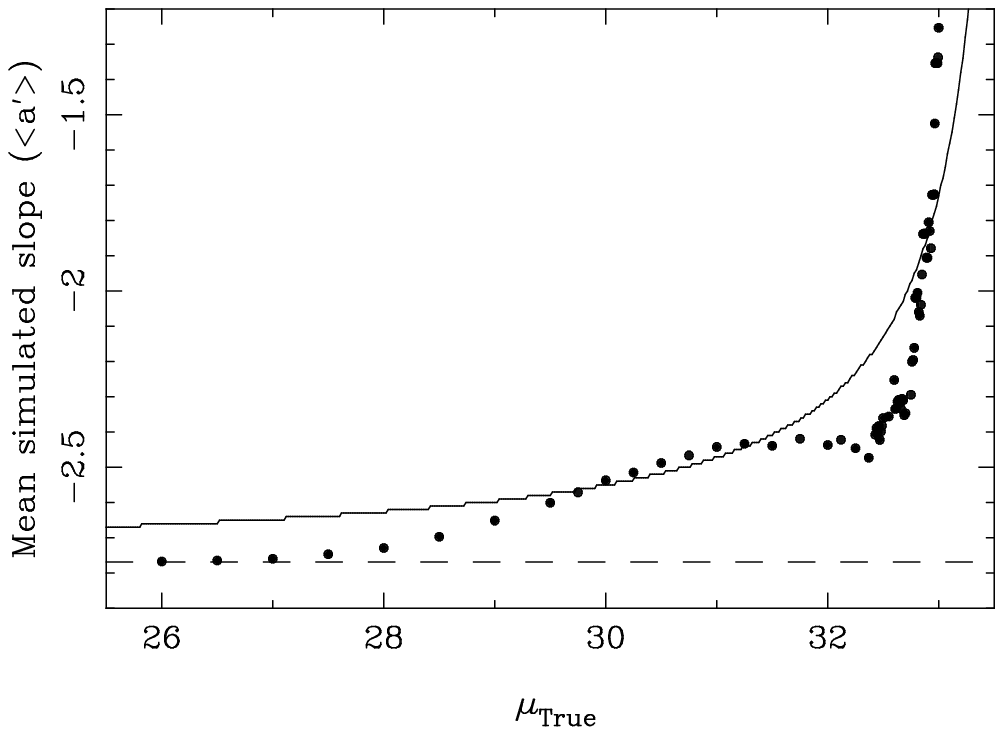]{ 
Simulation of the bias affecting the slope of the PL relation as a function
of the distance modulus. The plotted slope is the mean value of a set of 200
simulated galaxies having increasing distance moduli. 
The horizontal dotted line indicates the unbiased value.
The curve represents our model given by equation (\ref{slope}).
\label{fig8}}

\figcaption[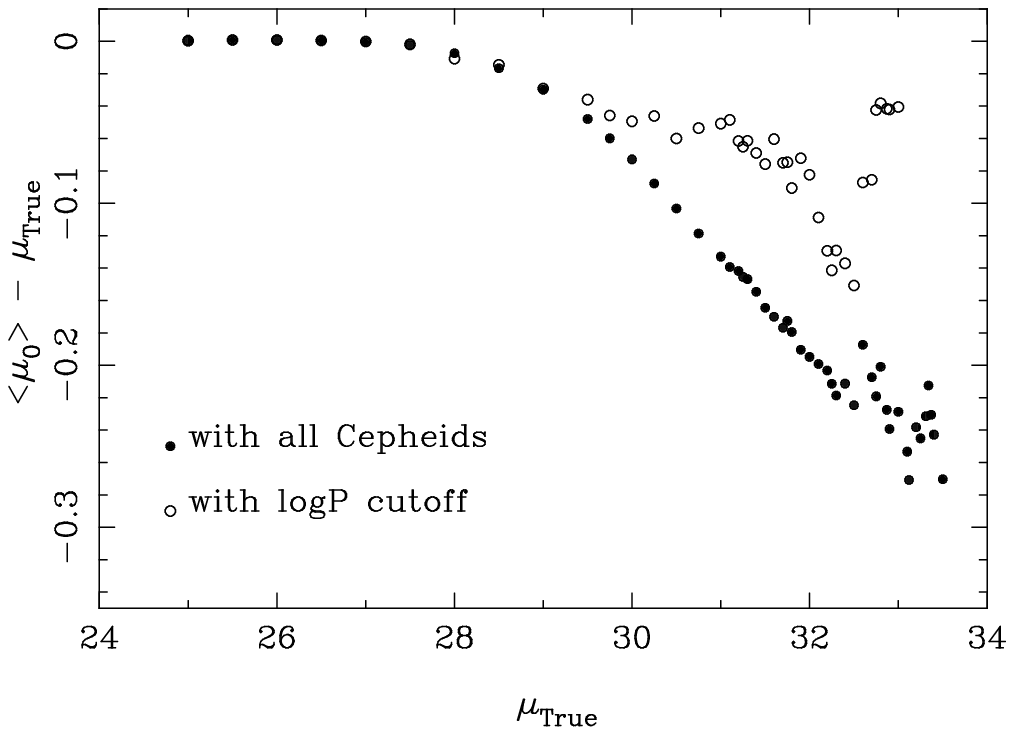]{
Simulation of the bias we made while computing the distance modulus of
a galaxy, because of incompleteness. Filled dots represent the mean
error on $\mu_{0} - \mu_{True}$ as a function of $\mu_{True}$, for a set
of 200 galaxies, when keeping all observed Cepheids in the calculation.
Open dots represent the same quantity when applying a cutoff in $\log P$
according to equation (\ref{cut}). One can see that this procedure significantly
reduces the error on the distance modulus.
\label{fig9}}

\clearpage
\plotone{pl4536.eps}
\clearpage
\plotone{pl4536i.eps}
\clearpage
\plotone{slope.eps}
\clearpage
\plotone{NGC4536.eps}
\clearpage
\plotone{simuplV.eps}
\clearpage
\plotone{simuplI.eps}
\clearpage
\plotone{SIMUL.eps}
\clearpage
\plotone{bias_slope.eps}
\clearpage
\plotone{bias_mu.eps}

\end{document}